# GrantMed: a new, international system for tracking grants and funding trends in the life sciences

Yuri Nikolsky[1,2]*, Roman Gurinovich[1,3]*, Oleg Kuryan[1,3], Aleksandr Pashuk[1,3], Alexej Scherbakov[1,3], Konstantin Romantsov[1], Leslie C. Jellen[4,5], Alex Zhavoronkov[5,6]*

[1] Infinity Sciences, Inc, 16192 Coastal Highway, Lewes, Delaware, County of Sussex, 19958, USA

[2] School of Systems Biology, George Mason University, VA, Manassas, 20110, USA

[3] Xpansa, Conzl OU, Mustamae Tee 5, Tallinn, 10616, Estonia

[4] Genetics, Genomics, and Informatics, University of Tennessee Health Science Center, Memphis, TN, 38163, USA

[5] Insilico Medicine, Inc, Johns Hopkins University, ETC, B310, Baltimore, MD, 21218, USA

[6] The Biogerontology Research Foundation, 2354 Chynoweth House, Trevissome Park, Blackwater, Truro, Cornwall TR4 8UN, UK

* To whom correspondence should be addressed. *Email: ynikolskiy@infinitysciences.com, *Email: roman.gurinovich@xpansa.com, *Email: alex@biogerontology.org

Present Address: The Biogerontology Research Foundation, 2354 Chynoweth House, Trevissome Park, Blackwater, Truro, Cornwall TR4 8UN, UK

## ABSTRACT

Despite the success of PubMed and other search engines in managing the massive volume of biomedical literature and the retrieval of individual publications, grant-related data remains scattered and relatively inaccessible. This is problematic, as project and funding data has significant analytical value and could be integral to publication retrieval. Here, we introduce GrantMed, a searchable international database of biomedical grants that integrates some 20 million publications with the nearly 1.4 million research projects and 650 billion dollars of funding that made them possible. For any given topic in the life sciences, Grantmed provides instantaneous visualization of the past 30 years of dollars spent and projects awarded, along with detailed individual project descriptions, funding amounts, and links to investigators, research organizations, and resulting publications. It summarizes trends in funding and publication rates for areas of interest and merges data from various national grant databases to create one international grant tracking system. This information will benefit the research community and funding entities alike. Users can view trends over time or current projects underway and use this information to navigate the decision-making process in moving forward. They can view projects prior to publication and records of previous projects. Convenient access to this data for analytical purposes will be beneficial in many ways, helping to prevent project overlap, reduce funding redundancy, identify areas of success, accelerate dissemination of ideas, and expose knowledge gaps in moving forward. It is our hope that this will be a central resource for international life sciences research communities and the funding organizations that support them, ultimately streamlining progress.

Database URL: www.grantmed.org

**INTRODUCTION**

Since its launch in 1996, PubMed has been the researcher's primary source for literature search and retrieval, hosting over 20 million publications[1] from over 5000 journals[2] and handling queries numbering in the millions each day.[3] Yet despite its success, the challenge of managing literature online continues.[4,5] A number of attempts at improving literature search and retrieval have been made, most notably Web of Science (http://wokinfo.com) and Google Scholar (http://scholar.google.com/) These have helped reel in publications widely scattered across disciplines and resources and pair them with other valuable data that may have otherwise been difficult to access. They offer users a single, simple platform and improvements in access and interface.

Despite the success of these efforts, one problem that we have seen go unaddressed is the failure to link publications to underlying grants, including research project and funding data, particularly at an international level. This data is indispensable. Tens of billions of dollars are spent annually in government-funded projects, primarily by NIH, NSF, and DARPA. The projects supported by these funds back the rising number of billions also spent annually in corporate research and development.[6] Over 1.5 trillion dollars are now locked into clinical trials in the pharmaceutical industry (https://clinicaltrials.gov). Ultimately, government funding leads to advances in biomedicine and new drugs in the corporate sector, and these in turn lead to new or improved therapies and diagnostics. New proposals and directions may be driven by investigators, but are ultimately steered by the funding agencies, universities, and research organizations that support them.

Data concerning where the funding goes and what projects succeed or fail has tremendous analytical value. Dollar-based decision-making within universities, research organizations, funding agencies, and laboratories paves the paths researchers ultimately pursue, and funding is inextricably linked to research directions, prioritization, and ultimately progress in science. A system that tracks grants, integrates them with publications, and provides descriptive summary statistics of funding and publication rates would enable the viewing of individual projects as well as the analysis of funding trends over the span of decades for a particular area of interest. This could have many applications, including identifying areas of funding drought or surplus and developing or considering new proposals.

Project descriptions are also valuable, particularly in light of the time delay between project awarding and publication. Individual research publications are most often funded within an umbrella project that spans years and involves a number of studies. At any given point, some of these may be published, some in preparation, and some proposed for a future date. Other publications or ideas may be discarded along the way as new data comes in. There are often years between the grant application and publication, followed by a time gap between the publication and "reduction to practice" of its results in a form of preclinical research in life science companies. Even more time and money are then spent on clinical trials. With all of this delay, it would be helpful to researchers, funding entities, and corporate organizations alike to view projects prior to publication.

**Introducing GrantMed**

To address this problem, we aimed to develop a new system wherein publications could be viewed in the context of grants and grants within the context of a thirty year grant landscape. We wanted it to be an international database. To promote international collaboration, we also wanted projects and publications linked to the funding agencies, investigators, and universities involved. Resulting would be one central platform for viewing not just the publication, but the entire publication process, with every contribution, from funding agency, university, and investigator to proposal, project, and funding amount to publication, visible and one click away.

Our goal in developing such a system was that it would help guide future directions and prevent project overlap, expose knowledge gaps and fruitful project directions. Furthermore, analysis of the current and past grant landscape could help rationally predict the directions and value of new technologies and pharmaceutical R&D pipelines over a 5-10 year horizon. For these reasons we have developed GrantMed. GrantMed is an attempt to systematize such knowledge with semantic categorizing, seamless data integration, and intuitive analytical interface.

**Existing project databases**

Two issues exist with the current system of tracking research projects and grant money: 1) project descriptions, grants, and investigators are not directly linked to publications in a centralized system, and 2) project tracking systems are national, not international. Thus, multiple national resources exist for tracking projects and grants; in the US, these include NIH RePORTER (report.nih.gov) and National Science Foundation (NSF; http:/www.nsf.gov/), in Canada, Canadian Institutes of Health Research Database (researchnet-recher-recherchenet.ca), and in Europe, Health Competence (health-competence.eu) and cordis.europa.eu. While these resources do have a number of useful features (in NIH RePORTER, for example, one can search investigators or grants to find funding amounts, view success rates, or visualize breakdowns of annual expenditures per research topic), they lack an international view of the data. To search projects or grants for a particular topic of the life sciences, users must search each of these sources individually. Moreover, the project and grant information is not always linked to the publications that result. The result is lost time and ultimately, since few have the time to search multiple sources regularly, a loss of valuable information.

METHODS

**Data sources**

A major goal in developing GrantMed was to build an international system that organizes life sciences research worldwide. Project data is sourced, therefore, by multiple national databases. At present, these include data from the US (NIH Reporter/Exporter and NSF), EU (Community Research and Development Information Service (CORDIS)), Australia (National Health and Medical Research Council (NHMRC)), Canada (Canadian Institutes of Health Research (CIHR)) and different international resources.

Other data include publications, researchers, universities, and university rankings and are also integrated into Grantmed.org from multiple sources. Publications are sourced by PubMed. Clinical trials data are from the International Clinical Trials Registry Platform (ICTRP) and the NIH service, ClinicalTrials.gov.

It should be emphasized that the data is being collected constantly and the database is updated daily. Updates are performed by extractors applications that incrementally add new records from the sources to the consolidated database. In the six months prior to the submission of this manuscript, approximately 300 scientists and collaborators, 200 projects, and 1100 publications were added daily. Presently, GrantMed consists of mainly North American and European research, including 3 million projects, 25 million publications , and a massive amount of other data on clinical trials, investigators, companies, etc. In the future, the scope of our work will include projects, publications, and funding data from Japan and the entire Asia region, as well, which should double number of records and daily growth. The long term purpose is to create one single global Life Sciences trends analytic software.

**Stages of development**

Development of the GrantMed project took place in three stages (Figure 1):

1.) Collecting data. In the first stage, data was consolidated from different sources into MongoDB and data extraction algorithms were developed to enable continuous data retrieval from various sources.

2.) Normalizing and transforming data. In the second stage, the data was cleared, structured and integrated into the MySQL database, to be used in the third stage. The architecture of the MySQL database and modules that export data from MongoDB to MySQL was also designed (Picture 2).

3.) Web application and reports generation development. At this stage, the frontend of GrantMed was developed, primarily targeting the portal for non-IT scientists at universities and research organizations. Completeness of content and user-friendly interface were the main goals. The frontend of GrantMed was implemented in Django CMS, a web platform written in Python. All charts were developed with Flot jQuery library (http://www.flotcharts.org/).

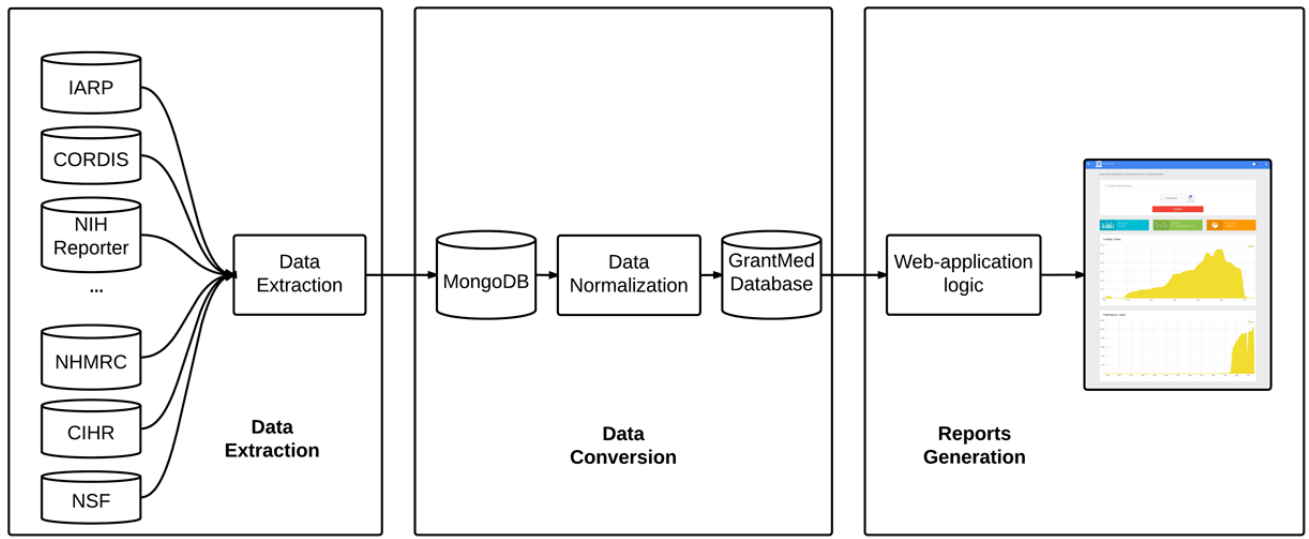

Figure 1. Structure of GrantMed project

**Content and Structure**

GrantMed is a one-page web-application with intuitive appearance. The search pane below the header (Fig.2) includes an input field for search terms and Google reCAPTCHA (https://www.google.com/recaptcha/intro/index.html) to protect the application from abuse and hacking while allowing legitimate users to pass through with ease.

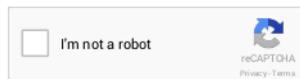
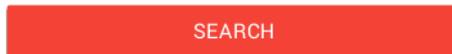

Figure 2. Search block of GrantMed

The next block consists of three similar elements: Total Projects, Total Funding, Total Publications (Fig. 3). On the main page, these elements display current information about the complete set of data in the GrantMed database. Once a search query is initiated by clicking on the search button, the application updates all content and to display the same summary of information pertaining to the specific research area searched.

NON-PROFIT BIOMEDICAL RESEARCH PROJECT SEARCH ENGINE

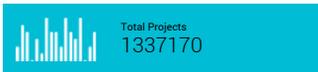
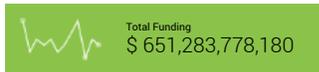
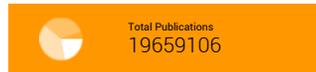

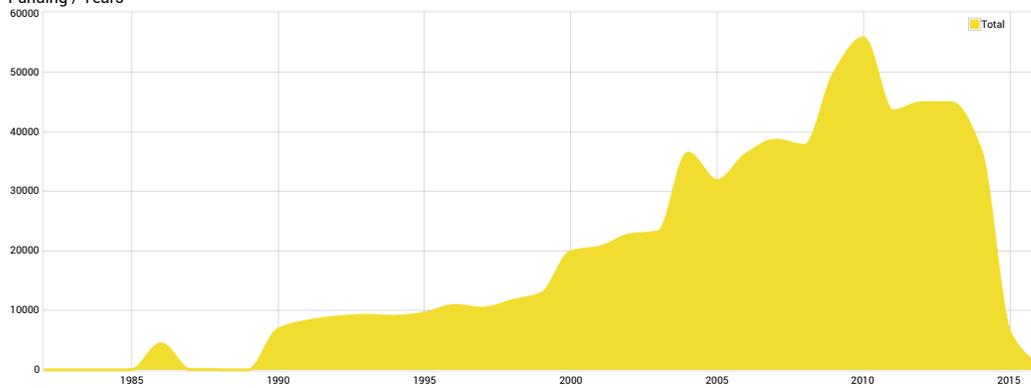

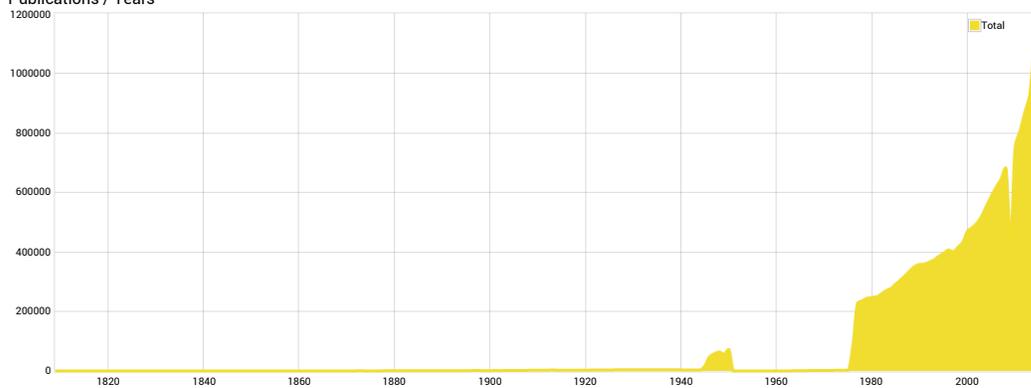

Figure 3. The Overview block

Below the overview information, there are two full-width charts (Fig. 4). The first one displays the distribution of funding by years; the second by publications. As in the previous case, the chart on the home page displays data summarizing the complete set of data on GrantMed. The GrantMed application updates the charts to reflect the search query once a search is initiated.

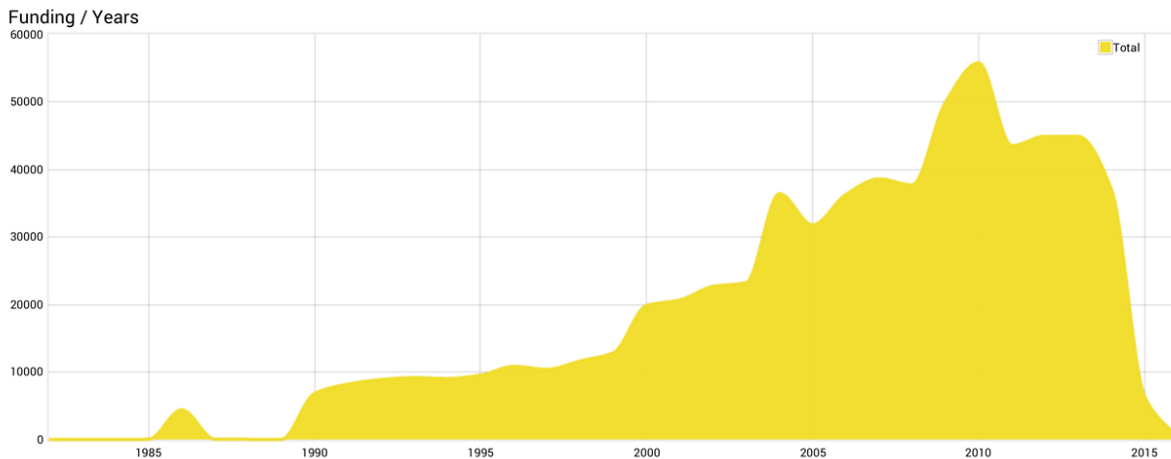

Figure 4. Example of Funding/Years chart

In addition to the summary charts for each search term, GrantMed also provides tables of data for statistical analysis. These include various key performance indicators (kpis), which help to answer the following questions:

1. What is the average cost of projects related to the search term? This can be used by funding entities and investigators to determine grant amounts and the number of collaborators possible.
2. How much research is performed in this area? This informs the user about area maturity.
3. How competitive is the area?
4. What is the social importance of the term under research?

Summary statistics for a search query include first year cited, number of publications, number of projects, year of first funding peak, year of first publication peak, year of first funding trend, and year of first publication trend.

All calculations are carried out by Xpansa Statistical Engine (https://xpansa.com/), a tool written on Python and based on SciPyPython library (http://www.scipy.org/) , a special ecosystem of open-source software for mathematics, science, and engineering.

The summary charts and statistics provide an instantaneous overview of the funding and publications surrounding the search query, but users may also be interested in individual projects related to the query. Below the summary charts, search results also include tables of descriptive project data. These are displayed as dynamic DataTables (https://www.datatables.net/), with options to sort and filter content. The last column of the Projects table (Fig. 5) opens a child DataTable with meta-data on related publications for the selected project (Fig.6).

Figure 5. An example of search results: projects

Figure 6. An example of child table with publications related to project

Publications are linked to Google Scholar in the last column of the Publications table (Figure 7). Column headers allow users to sort results by different columns.

Figure 7. An example of search results: publications

Most operations between the frontend and the backend are performed with AJAX (Asynchronous JavaScript and XML). AJAX enables quick and easy interactions between the user and website, as pages are not reloaded for the displayed content. AJAX also reduces traffic between the client machine and the application server. At this time, the application uses the MySQL full-text search feature as a default search engine. While this is sufficient for the current version of GrantMed, we ultimately plan to switch to the Sphinx search engine, which is fully open source and one of the best search engines available. With a rapid influx of data, Sphinx can provide a faster and more customizable search. We also plan adding filters for different meta-data parameters: project types, publication types, resources, organization, etc. This set will cover most use cases and enable scientists to quickly access necessary information.

**Use Case**

To provide a case of system application, we selected a high impact protein, programmed death 1 receptor PD-1, and explored the related projects and publications data. PD-1 and its ligands are promising targets for checkpoint inhibitors in anti-cancer interventions.[7]

A search querying PD-1 reveals a total of 390 related projects, 2628 publications, and over $180,000 in funding. The first related publications dating back to the 1980s, the most significant first work being a publication in 1992 by Ishida et al. revealing the induction of PD-1 upon programmed cell death.[7] From the summary charts, it is apparent that PD-1 research rose in the second half of the 2000s. According to the publication data, this trend was initiated by a publication by Racke and Stewart (2002) on T cell costimulation in autoimmune disease.[8] There was also a notable rise of publications in 2007 and funding in 2008, with a decrease in publications in 2014 and correlation in topic development cycle.

From these examples, one can see that the cycle from novelty to practical application to worldwide concept development is around 19 years in average. For PD-1, this long and winding road from discovery to clinical application is illustrated in the review by Okasaki and Honjo (2007).[9] This highlights several issues in science, including the need for:

1. Global interdisciplinary scientific collaboration to lower knowledge transfer delay.

2. *In silico* modeling to accelerate pace of of studying drugs and substances in model organisms and ultimately humans.

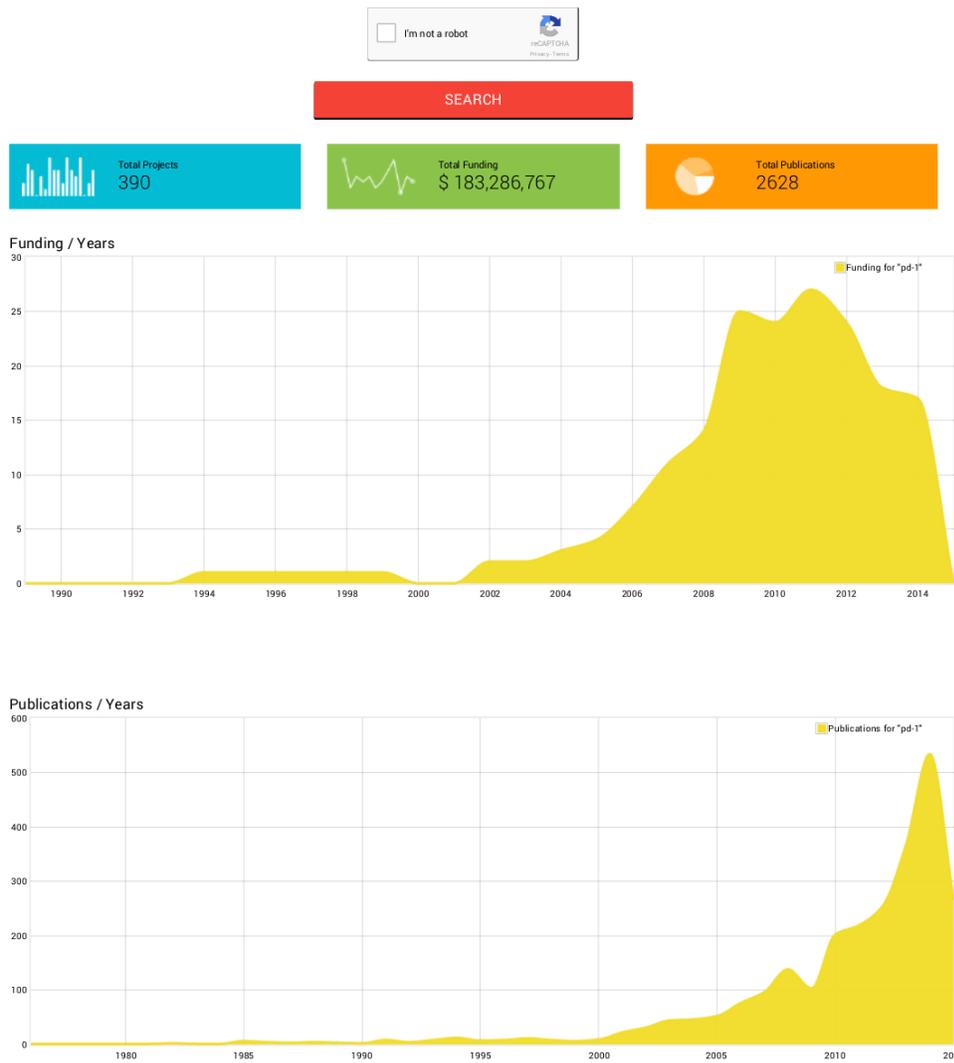

Figure 8. Summary trend valuation case for PD-1.

| Term | YFC | N, pub | N, prj | $, funding | First Pub Peak | First Prj Peak | Pub trend | Project trend |
|---|---|---|---|---|---|---|---|---|
| PD-1 | 1992 | 3066 | 768 | 336M | 1996 | 2002 | 2008 | 2007 |
| PD-L1 | 1992 | 1297 | 300 | 152M | 1996 | 2002 | 2008 | 2007 |
| CTLA-4 | 1989 | 3539 | 1069 | 372M | 1995 | 1997 | 2000 | 2001 |

| Checkpoint Inhibitor | 1999 | 222 | 34 | 18M | 2003 | 2004 | 2012 | 2014 |

CONCLUSION

Ultimately, every publication traces back to a research project awarded and dollars spent.  The record of past and present projects awarded and dollars spent should thus be integral to the decision-making process in moving forward, for researchers and funding entities alike.  Yet, these two information-rich data sources remain relatively inaccessible.  This is particularly true at the international level.  To integrate biomedical literature with underlying research projects awarded and improve funding visibility, we have created Grantmed.org.  Grantmed is an international organizational system for tracking grants in the life sciences.  It is a searchable database that integrates publications with research projects, research teams, and funding amounts, allowing researchers or funding agencies to view trends over time or current projects underway in a given research area.  It will aid in navigating the decision-making process when choosing new avenues to fund or explore, helping to prevent project overlap, reduce funding redundancy, identify areas of success, accelerate dissemination of ideas, and expose knowledge gaps in moving forward.  It is our hope that this will be a valuable resource for international life sciences research communities and the funding organizations that support them.